\documentclass[sigconf,screen]{acmart}

\renewcommand\footnotetextcopyrightpermission[1]{}
\copyrightyear{2024} 
\acmYear{2024} 
\setcopyright{rightsretained} 
\acmConference[ICSE-Companion '24]{2024 IEEE/ACM 46th International Conference on Software Engineering: Companion Proceedings}{April 14--20, 2024}{Lisbon, Portugal}
\acmBooktitle{2024 IEEE/ACM 46th International Conference on Software Engineering: Companion Proceedings (ICSE-Companion '24), April 14--20, 2024, Lisbon, Portugal}\acmDOI{10.1145/3639478.3643518}
\acmISBN{979-8-4007-0502-1/24/04}

\usepackage{subcaption}
\usepackage{multirow}
\usepackage{colortbl}
\usepackage{svg}
\usepackage{adjustbox}
\usepackage[utf8]{inputenc}
\usepackage{amsmath}
\usepackage{algorithm}
\usepackage[noend]{algpseudocode}
\usepackage{macros}
\usepackage{tabularx}
\usepackage{graphicx}

\newlength{\textfloatsepsave}
\title{Improving Program Debloating with 1-DU Chain Minimality}

\ccsdesc[500]{Security and privacy~Software security engineering}

\author{Myeongsoo Kim}
\affiliation{
  \institution{Georgia Institute of Technology}
  \city{Atlanta}
  \state{Georgia}
  \country{USA}
}
\email{mkim754@gatech.edu}

\author{Santosh Pande}
\affiliation{
  \institution{Georgia Institute of Technology}
  \city{Atlanta}
  \state{Georgia}
  \country{USA}
}
\email{santosh@cc.gatech.edu}

\author{Alessandro Orso}
\affiliation{
  \institution{Georgia Institute of Technology}
  \city{Atlanta}
  \state{Georgia}
  \country{USA}
}
\email{orso@cc.gatech.edu}

\begin{document}

\begin{abstract}
Modern software often struggles with bloat, leading to increased memory consumption and security vulnerabilities from unused code. In response, various program debloating techniques have been developed, typically utilizing test cases that represent functionalities users want to retain. These methods range from aggressive approaches, which prioritize maximal code reduction but may overfit to test cases and potentially reintroduce past security issues, to conservative strategies that aim to preserve all influenced code, often at the expense of less effective bloat reduction and security improvement. In this research, we present RLDebloatDU, an innovative debloating technique that employs 1-DU chain minimality within abstract syntax trees. Our approach maintains essential program data dependencies, striking a balance between aggressive code reduction and the preservation of program semantics. We evaluated RLDebloatDU on ten Linux kernel programs, comparing its performance with two leading debloating techniques: Chisel, known for its aggressive debloating approach, and Razor, recognized for its conservative strategy. RLDebloatDU significantly lowers the incidence of Common Vulnerabilities and Exposures (CVEs) and improves soundness compared to both, highlighting its efficacy in reducing security issues without reintroducing resolved security issues.
\end{abstract}

\maketitle

\section{Introduction}

The growing complexity and bloat of software systems have become a challenging issue in recent years, leading to increased memory consumption and security vulnerabilities originating from unused code~\cite{hibbs2009art}. As developers strive to meet the demand for feature-rich software, they frequently incorporate new functionalities, inadvertently expanding the codebase size and intricacy, leading to bloat~\cite{holzmann2015code}. This trend exacerbates the challenges associated with managing software bloat, negatively impacting system performance and increasing security vulnerabilities. 

In a response to this, numerous debloating approaches aim to reduce program size as much as possible, yet achieving a globally minimal program for real-world applications is nearly impossible. This challenge has led to the adoption of 1-minimality as a practical concept in debloating~\cite{zeller2002simplifying}. A program is considered 1-minimal if removing any single element does not maintain its desired properties, ensuring compactness while preserving functionality. Notably, Chisel~\cite{heo2018effective} focus on eliminating unnecessary code while maintaining syntactic correctness, with employing 1-tree minimality~\cite{misherghi2006hdd} and integrating reinforcement learning to streamline the process. However, these aggressive strategies may inadvertently remove essential security patches, reintroducing vulnerabilities~\cite{qian2019razor}. 


Conversely, conservative methods like Razor~\cite{qian2019razor} prioritize retaining all relevant code to prevent security lapses. Razor employs control-flow-based heuristics to infer related code by examining paths similar or closely related to the executed ones. However, since Razor tends to retain all code influenced by these heuristics, it can result in bloated software and may fail to eliminate some critical CVEs that more aggressive methods can address.



\begin{figure}[t]
  \centering
  \includegraphics[width=0.48\textwidth]{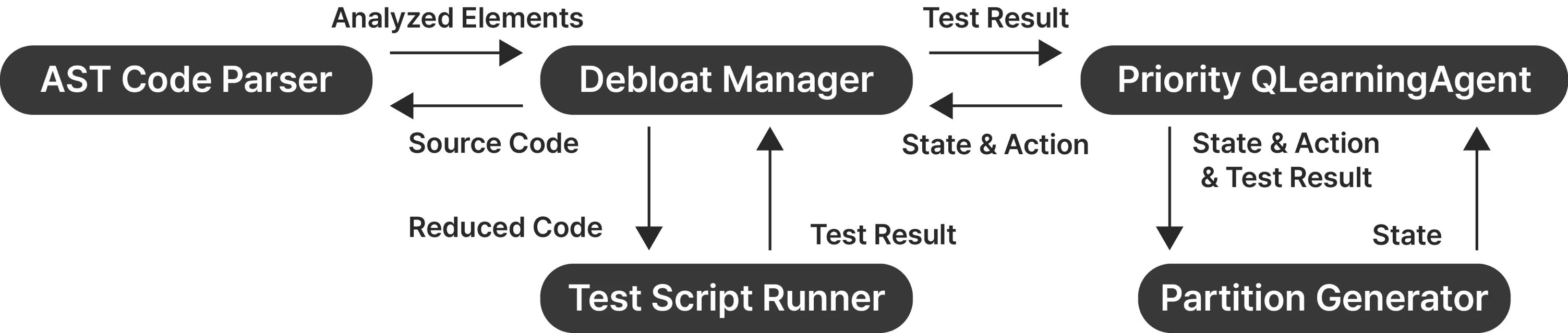}
  \caption{A diagram of RLDebloatDU's Architecture.}
  \label{fig:architecture}
\end{figure}

In this paper, we introduce RLDebloatDU, a debloating technique that achieves 1-DU Chain minimality with reinforcement learning-guided delta debugging. A DU chain represents the sequence of definitions and uses of a variable within a program, ensuring each definition reaches a use without being redefined. Our novel approach utilizes 1-DU chain minimality to preserve data semantics by {\it preserving all needed DU chains}, striving to offer effectiveness and security benefits in program debloating. Unlike Razor, which employs control-flow-based heuristics which retains all potentially influenced code, RLDebloatDU's focus on 1-DU chain minimality ensures a more precise debloating. By integrating reinforcement learning, RLDebloatDU accelerates the delta debugging process, enabling more efficient exploration of the search space.

\begin{table*}[t]
\centering
\caption{Gadget counts for reduced versions generated by Chisel, RLDebloatDU, and Razor}
\vspace{-10pt}
\label{gadgets}
\begin{tabularx}{\textwidth}{|l|*{15}{>{\centering\arraybackslash}X|}}
\hline
 {Program}& \multicolumn{5}{c|}{RLDebloatDU} & \multicolumn{5}{c|}{Chisel} & \multicolumn{5}{c|}{Razor} \\
\cline{2-16}
 & Total & ROP & JOP & COP & SP & Total & ROP & JOP & COP & SP & Total & ROP & JOP & COP & SP \\
\hline
Bzip2-1.0.5 & 34.2\% & 28.3\% & 85.7\% & 33\% & 100\% & 42.4\% & 33.7\% & 66.7\% & 51.9\% & 100\% & 3.8\% & 2.1\% & 47.6\% & -2.8\% & 50\% \\
Chown-8.2 & 73\% & 62.3\% & 87.5\% & 97.2\% & 100\% & 72.8\% & 62\% & 87.5\% & 97.2\% & 100\% & 44\% & 31.3\% & 70.8\% & 69.4\% & 88.9\% \\
Date-8.21 & 42.3\% & 46.5\% & 29.6\% & 11.8\% & 85.7\% & 50.3\% & 46.5\% & 77.8\% & 47.1\% & 100\% & 24.4\% & 30.1\% & 14.8\% & -24\% & 85.7\% \\
Grep-2.19 & 73.4\% & 55.3\% & 96.5\% & 84.6\% & 100\% & 69.7\% & 47.9\% & 98.1\% & 82.1\% & 100\% & 59.5\% & 35.6\% & 97.5\% & 55.3\% & 92.3\% \\
Gzip-1.2.4 & 57.5\% & 41.2\% & 55.6\% & 88.8\% & 100\% & 63.3\% & 51\% & 83.3\% & 86.7\% & 42.9\% & 62.4\% & 49.5\% & 55.6\% & 88.8\% & 85.7\% \\
Mkdir-5.2.1 & 54.4\% & 52.8\% & 66.7\% & 50\% & 100\% & 52.3\% & 50.4\% & 66.7\% & 50\% & 100\% & 10.1\% & 7.9\% & 60\% & -67\% & 0\% \\
Rm-8.4 & 70.7\% & 66.9\% & 66.7\% & 93.5\% & 100\% & 69.4\% & 66.2\% & 33.3\% & 93.5\% & 100\% & 25\% & 24.1\% & 33.3\% & 21.7\% & 83.3\% \\
Sort-8.16 &	76.1\% &	70.5\% &	93.9\% &	 94.3\% &	100\%	 & 78.8\% &	73.9\% &	93.9\% &	95.7\% &	100\% & 31.7\% &	31.9\% &	55.1\% &	21.4\% &	-50\% \\
Tar-1.14 &	82.8\% &	71.2\% &	97.4\% &	95.3\% &	100\% & 84.9\% &	74.5\% &	98.5\% &	95.7\% &	100\% &	69.9\% &	57.1\% &	89.2\% &	82.5\% &	 66.7\% \\
Uniq-8.16 &	69.5\% &	67.3\% &	78.6\% &	77.3\% &	100\%	 & 68.7\% &	66.8\% &	71.4\% &	77.3\% &	100\% & 36.2\% &	28.7\% &	92.9\% &	59.1\% &	80\% \\
\hline
Average &	63.39\% &	56.23\% &	75.82\% &	72.58\% &	98.57\% & 65.26\% &	57.29\% &	77.72\% &	77.72\% &	94.29\% &	36.7\% &	29.83\% &	61.68\% &	30.52\% &	58.26\% \\
\hline
\end{tabularx}
\vspace{-12pt}
\end{table*}

We evaluate our method on ten Linux kernel programs, using test cases provided by Chisel, and compare our results to state-of-the-art approaches, Chisel and Razor. Our assessment focused on key metrics including binary size, CVE code, soundness, and various types of gadgets, such as ROP (Return-Oriented Programming), JOP (Jump-Oriented Programming), COP (Call-Oriented Programming), and special purpose gadgets~\cite{brown2019less}. Preliminary results indicate that RLDebloatDU demonstrates significant advancements in program debloating by effectively balancing code reduction with enhanced security, without reintroducing previously resolved vulnerabilities. 

\section{Approach and Evaluation}
\label{sec:our_approach}

RLDebloatDU integrates key components systematically to reduce software bloat while ensuring functionality retention, as illustrated in Figure~\ref{fig:architecture}. The process begins with the AST Code Parser analyzing the codebase using Abstract Syntax Trees to identify debloatable segments. The PriorityQLearningAgent then employs Q-Learning to determine the most effective debloating policy.
This strategy is orchestrated by the DebloatManager, which coordinates the debloating process and ensures functional integrity. The Test Script Runner executes test cases on the debloated code to verify its continued functionality. Complementing this, the Delta Debugging based Partition Generator efficiently generates code partitions for debloating testing. RLDebloatDU's operation involves an iterative loop: analyzing the code, updating Q-values based on test results, and generating progressively leaner code versions. This cycle targets functions, global and local variables, including their Def-Use chains, and persists until no further code size reduction is possible.

To evaluate RLDebloatDU, we conducted tests on ten Linux kernel programs from Chisel's benchmark, using a Google Cloud E2 machine with Ubuntu 20.04. The goal was to benchmark RLDebloatDU against existing debloating tools, namely Chisel and Razor. Our findings showed RLDebloatDU's exceptional ability to enhance security by significantly reducing CVEs. The technique managed to lower the total number of CVEs from 10 in the original code to just 4, outperforming Chisel, which reduced them to 7, and Razor, which achieved a reduction to 6. Notably, RLDebloatDU was as effective as Chisel in removing CVEs, without reintroducing previously resolved security vulnerabilities.

In addition to its notable security benefits, RLDebloatDU also excelled in reducing gadgets, a crucial factor in minimizing a program's attack surface. As highlighted in Table~\ref{gadgets}, RLDebloatDU achieved an average gadget reduction of 63.39\%, significantly better than Razor's 36.7\% and closely comparable to Chisel's 65.26\%. Further, in assessing the binary reduction rate for executable segments, RLDebloatDU achieved an average reduction of 83.57\%, slightly below Chisel's 89.71\% but considerably higher than Razor's 54.37\%. These results underscore RLDebloatDU's balanced approach in enhancing security while efficiently reducing program size.

Lastly, we conducted an additional experiment using Razor's test cases, which included 299 tests for 10 Linux programs. We found that RLDebloatDU crashed in one program - Tar, while Chisel crashed in four programs (Bzip2, Grep, Gzip, and Tar), and Razor also crashed in four programs (Chown, Date, Rm, and Tar). These results demonstrate that RLDebloatDU is more sound than both Chisel and Razor.

\section{Conclusion}\label{sec:conclusion}
We introduced RLDebloatDU, a novel reinforcement learning-based software debloating approach that efficiently reduces a program's attack surface while preserving its functionality. We made our artifact publically available for future research and result reproduction~\cite{artifact}. Future work will involve exploring a variety of learning algorithms and automatic feature identification, in addition to comparisons with more tools~\cite{xin2020program, xin2020subdomain}.

\begin{acks}
	  This work was partially supported by 
	  NSF, under grant CCF-0725202,
	  DOE, under contract DE-FOA-0002460,
	  and gifts from Facebook, Google, IBM Research, and Microsoft Research.
\end{acks}

\balance
\bibliographystyle{ACM-Reference-Format}
\bibliography{bib}

\end{document}